\documentclass{andp2012}
\usepackage[english]{babel}
\title{Why we need an electron-ion collider}
\author[F. Author]{Raju Venugopalan\inst{1}\footnote{\quad E-mail:~\textsf{raju@bnl.gov}}}
\address[1]{Physics Department, Brookhaven National Laboratory, Upton, NY 11973, USA}
\shortauthors{R. Venugopalan}
\begin{abstract}
 We present a brief argument making the science case for an electron-ion collider.
\end{abstract}
\shortabstract
\begin{document}
\maketitle

\section{The puzzling structure of the hadron}
 The distinction between fundamental constituents and the emergent phenomena they generate is perhaps most opaque in  
Quantum Chromodynamics (QCD), the fundamental theory of nature's strong force. For instance, while the fundamental constituents that make up the proton and neutron building blocks of nuclear physics are massless gluons and the nearly massless up and down quarks, these contribute only a small fraction of the proton's mass. The rest arises from complex quark-gluon dynamics. Central to this complex dynamics are the spontaneous breaking of the chiral symmetry of the QCD Lagrangian (with massless quarks) and color confinement. The latter enforces the color charge of quarks and gluons to be confined on distance scales of less than a femtometer, with strongly interacting matter at longer distances organized into color singlet hadron states. These features of QCD,  ill-understood despite much progress in numerical lattice techniques, stand in the way of a dynamical understanding of the fundamental structure of hadron matter. 

As a simple example, consider trying to explain to a non-expert what the proton looks like.  Radically different pictures  of hadrons  exist at low energies relative to those at high energies. In the former case, a description of hadrons in terms of constituent quarks (and refinements thereof) describes many features of hadron spectra. In the latter, the parton model description in terms of current quarks and gluons provides much of our intuition in describing high energy scattering. How to relate the two pictures remains a mystery. Another obvious question whose answer remains a puzzle is how the quark and gluon constituents of the proton make up its spin. Naively one expects this to be the vector sum of the constituent quark spins. However as demonstrated by the Spin Muon Collaboration (SMC), only $\sim 30\%$ of the proton's spin is from the spin carried by quarks and anti-quarks. The rest arise from a complex interplay of gluon spin, and the orbital motions of quarks and gluons. How these are apportioned, and what that tell us about momentum and spatial correlations in the proton, are outstanding puzzles. 

The potential importance of orbital motion suggests that one-dimensional Bjorken $x$ distributions are insufficient for understanding how quarks and gluons generate the proton's spin. A richer tomographic picture is required which quantifies  correlations between the spin of the proton and the distributions of polarized and unpolarized quarks and gluons in transverse momentum and impact parameter space. This information will enable one to construct three dimensional dynamical images of where quarks and gluons clump together in the proton, how they generate orbital motion and how they interact with polarized and unpolarized colored probes. Only preliminary steps have been taken in this direction.

Just as the proton's spin confounds the constituent picture of hadrons, the parton model picture predicts what appears to be a runaway cascade of gluons and sea quarks; successive rungs of this cascade carry smaller and smaller fractions $x$ of the proton's momentum. The Froissart-Martin theorem (based on unitarity, the absence of massless particles in the physical spectrum and reasonable assumptions about the functional form of the energy dependence of high energy cross-sections) tells us that the growth of small $x$ cross-sections cannot continue at the presently measured rate in deeply inelastic scattering (DIS) experiments off the proton at HERA. How does one understand dynamically how this growth tapers off? 

An intriguing idea to resolve this puzzle is that even though gluons are bosons, net repulsive recombination and screening processes generated by their many-body self-interactions work against a Markovian predilection for them to further split into gluons with smaller momentum fractions. These competing effects dynamically generate a scale that, with increasing energy, screens color charges at smaller and smaller distances. If true, it  would suggest that scattering processes in the high energy ``Regge-Gribov" limit of QCD are controlled by this emergent hard scale. A remarkable consequence is that, due to the asymptotic freedom property of QCD, important aspects of the corresponding many-body QCD dynamics can be computed systematically using effective field theory techniques.   

Quark and gluon parton distribution functions (PDFs) and correlations in nuclei are largely terra incognita. Ever since the first DIS experiments off fixed nuclear targets, it is known that nuclear quark and gluon distributions are not simple superpositions of nucleon quark and gluon distributions but show large deviations from these. How are the dynamics of quarks and gluons in nuclei different from those in vacuum? How does the transition to hadron degrees of freedom occur? Likewise, the multiple scattering of colored probes inside nuclei, their radiation and subsequent hadronization in matter, is poorly understood. Since the distinction between the fundamental constituents in QCD and their emergent degrees of freedom is thin, detailed QCD studies in nuclei may shed light not only on interesting and unexpected features of the quark-gluon structure of nuclei and their interactions, but perhaps on fundamental features of the underlying theory as well. For a discussion of these topics for a broader audience, see \cite{EUV}. 

\section{Resolving hadron puzzles with an EIC}

The exemplary DIS collider was HERA which provided remarkably precise information on parton distributions in the proton, vastly extending the phase space in Bjorken $x$ and $Q^2$. Since gluon fusion is the dominant channel for Higgs production, this data from HERA played a key role in the Higgs discovery at the LHC and continues to be crucial in searches for new physics beyond the standard model.  

The Electron-Ion Collider (EIC) proposal in the US, documented in a number of reports\cite{Boer:2011fh,Accardi:2012qut,Brodsky:2015aia}, will not have the energy reach of HERA;  the maximum electron-proton energy considered in present designs is $\sqrt{s}\sim 166$ GeV, approximately half that of HERA. However the EIC will be the world's first polarized electron-ion collider, it will be the world's first electron-nucleus collider (with a maximum $\sqrt{s}\sim 90$ GeV/nucleon) and it will have a luminosity of $10^2$ to $10^3$ times that of HERA. Further, it is anticipated that the detector coverage will be more hermetic than the HERA detectors, providing in particular a wide coverage in rapidity. The LHeC proposal at CERN has electron-proton $\sqrt{s}\sim 1.3-2$ TeV, approximately four to six times that of HERA and somewhat lower energy per nucleon electron-ion collisions as well. It will however not have polarized hadron beams~\cite{AbelleiraFernandez:2012cc}. Unless otherwise stated, our discussion will focus on the US EIC effort.  

The novel capabilities of the EIC machine (and potentially improved detectors) will allow QCD studies of fundamental aspects of hadron structure that are unique and will be unmatched by any other machine in versatility and precision. As we shall also discuss, the EIC will build on the scientific discoveries at RHIC, Jlab and the LHC, and provide essential information necessary to solve numerous puzzles raised by experiments at these machines regarding the sub-femtoscale structure of matter.  Here we shall provide a brief outline of some of the measurements that are only feasible with a polarized EIC, as discussed in detail in the EIC whitepaper~\cite{Accardi:2012qut} . 

\subsection{\em Nailing down the quark and gluon spin content of the proton}
The EIC will extract the polarized structure function $g_1(x,Q^2)$ to approximately two orders of magnitude lower values of $x$ for fixed $Q^2$ (and conversely in $Q^2$ for fixed $x$). This expansion in the $x-Q^2$ range and precision of the data, and the large lever arm in $Q^2$, will enable extraction of the nucleon gluon spin distribution $x\Delta g(x)$ to much lower $x$, with much smaller uncertainties, than feasible at RHIC. A significant achievement of the RHIC polarized proton collision experiments, was the demonstration of a non-zero $\Delta G = \int_{x_{\rm min}}^1 \Delta g(x)=0.2\pm 0.06\pm 0.07$ at $Q^2=10$ GeV$^2$, for $x_{\rm min}=0.05$. While the result of a global analysis of world data~\cite{deFlorian:2014yva}, the dominant contribution to this result is from RHIC. Nevertheless, when $x_{\rm min} = 0.001$, well below the $x$ values accessible at RHIC, the same analysis shows that 
the uncertainty in $\Delta G$ is large; its value is not constrained even to be positive. 

This is where the EIC will have a huge impact. Global analyses with mock EIC data suggest that both the quark spin $\Delta \Sigma$ and $\Delta G$ will be constrained to very high accuracy. With these constrained, the spin sum rule will deliver a quantitative determination of the orbital contributions to the proton's spin. Through semi-inclusive measurements of pion and kaon distributions, the EIC will also enable extraction of the polarized 
light flavor $\Delta u$, $\Delta d$ and strange quark $\Delta s$ (and respective anti-quark) polarized distributions with much greater precision and reach than extant studies. Additional insights into the helicity structure of the 
proton will be feasible with DIS of polarized Deuteron or Helium-3 beams allowing extraction of the neutron structure function $g_1^n (x,Q^2)$ (and therefore the Bjorken sum rule to greater precision than the current $10\%$ accuracy). Further, the high $Q^2$ of EIC will enable first measurements of polarized DIS in the regime governed primarily by virtual electroweak boson exchange. An example of such a measurement are the $g_{1,5}^{W^-}$ structure functions; these permit fragmentation function independent extractions of combinations of flavor dependent quark helicity distributions. 

Put together, data from an EIC would produce an enormous wealth of precise data on the helicity structure of the hadron. This enhances the likelihood of dramatic changes to our understanding of how the spin  of the proton is constituted, thereby providing a unique dynamical window into its quark-gluon sub-structure. 

\subsection{\em Three dimensional images of the internal structure of hadrons}

The above discussion concerned quark and gluon distributions that are one-dimensional distributions (in the momentum fraction $x$) evaluated at the resolution scale $Q^2$ of the DIS probe. However a deeper understanding of parton structure can be obtained by measuring their distributions in the 2+1-dimensional space of transverse momentum ${\bf k}_T$ and $x$ or of the impact parameter ${\bf b}_T$ and $x$. The former are represented by transverse momentum dependent distributions (TMDs) that can be measured in semi-inclusive DIS (SIDIS). The latter are determined from taking a Fourier transform of generalized parton distributions (GPDs) that are measured in exclusive DIS. Both of these types of distributions can be understood as arising from an underlying parton Wigner distribution in transverse momentum and impact parameter space~\cite{Belitsky:2003nz}. 

When the proton is polarized, the TMDs and GPDs are sensitive to spin-orbit correlations and  provide complementary pictures of these.  As a particular example, in a transversely polarized target, the unpolarized target quarks exhibit an anisotropy in their transverse momenta, characterized by a TMD called the Sivers function. This anisotropic distribution is extracted in SIDIS measurements where  the target is transversely polarized. The Sivers effect can be understood in a semi-classical picture as resulting from the final state interactions of the active up or down quark (that is hit by the virtual photon) with spectator partons that are spatially distorted in the polarized proton~\cite{Burkardt:2002hr}. The transverse spatial distortions in the polarized proton (which correspond to a flavor dipole moment) thus generate an anisotropy in transverse momentum space; evidence for this {\it chromodynamic lensing} picture has been adduced from HERMES and COMPASS data. In addition to the Sivers function, a number of other interesting spin-orbit correlations have been identified. These provide fundamental insight into the universality of initial and final state interactions in QCD and more generally into color flow and hadronization in the presence of spin-dependent interactions. 

The transverse momentum structure of polarized protons in the valence region will be thoroughly studied with Jlab 12 GeV. Where EIC would have the maximal impact is in the extension of such studies to the regime dominated by sea quarks and gluons;  gluon TMDs will be accessed for the first time. These may impact, for instance, more differential studies of Higgs production via the gluon fusion process. Further, the much larger $Q^2$ lever arm at EIC will allow clean separation of the leading twist TMDs (that have a clean interpretation in the Bjorken limit) from higher twist effects, enabling the study of both. The latter encode quark-gluon correlations, providing essential information on the many-body structure of the nucleon. TMD studies with mock EIC data show clearly these distributions can be measured with high precision, thereby permitting crisp extraction of 2+1-D images in transverse momentum space over a wide $x$ range. 

Likewise, big improvements in 2+1-D images in $x$-${\bf b}_T$ space from GPDs can be anticipated with exclusive measurements at EIC. For polarized protons, such measurements will extend to the gluon dominated regime for the first time well beyond the regime explored by HERMES, COMPASS and Jlab 6 and 12 GeV experiments. Even for unpolarized protons, the GPD measurements at HERA were challenging due to limited statistics. Understanding how quark and gluon distributions extend out in impact parameter in a controlled fashion is of great interest because one anticipates the transition from these degrees of freedom to meson degrees of freedom to occur in the tails of this distribution; studying this transition with precision, and a wide lever arm, has great potential for uncovering novel aspects of the dynamics of chiral symmetry breaking and confinement. Such studies thus far are in their infancy. A noteworthy point is that while some information on TMDs may be gleaned from hadron-hadron scattering experiments, the same is extremely challenging for GPDs, due to the ubiquity of soft color exchanges with spectators. Thus measurements of spatial gluon distributions in polarized hadrons and in nuclei will be absolutely unique measurements that are only feasible with an EIC. 

Finally, we note that there are ongoing theoretical efforts to compute both TMDs and GPDs on the lattice. These developments, combined with a deeper understanding of QCD factorization and evolution, and potentially with more sophisticated models and effective field theories, promise a much richer picture of the dynamical structure of hadrons. We will finally be able to proceed beyond the highly reductive one dimensional distributions that have thus far informed our understanding.

\subsection{\em Exploring the terra incognita of the nuclear quark-gluon landscape}
At low energies and large distances, nuclei are well described in terms of hadron degrees of freedom. However when these restrictions are relaxed, quark-gluon degrees of freedom become increasingly manifest. How does this transition take place? The EMC effect and shadowing, as well as intriguing information from Jlab linking large $x >1$ distributions with short range nucleon-nucleon correlations, suggest that the quark-gluon landscape of nuclei is complex and full of surprises. As the first DIS nuclear collider, the EIC will cover this vast landscape in atomic number $A$, and in $x$ and $Q^2$, with unprecedented versatility and precision. What the extended lever arm in $A$ provides is not an opportunity for stamp collecting, attractive as it may be to some, but the ability to decipher patterns that may or may not scale with the size of the nuclear medium.  

Remarkably little is known about even the most inclusive quantities, the structure functions $F_2$ and $F_L$, especially at small $x$. The fixed target data below $x=0.01$ are relatively low statistics data and at low $Q^2$. As a consequence, the sea and gluon distributions are ill constrained and the uncertainties are large, especially at low $Q^2$ scales. In particular, the lack of a  wide lever arm in $Q^2$ has inhibited the extraction of gluon distributions, the most efficient extraction of which are through the Bjorken scaling violations of $F_2$. Before the EIC turns on, considerable information on nuclear PDFs may become available from proton-nucleus collisions at RHIC and the LHC; we note too that ultraperipheral A+A collision data is of interest in this regard. However since the extraction of PDFs in hadron collision experiments involves additional convolutions, and fragmentation function uncertainties, these will be most reliable at large $Q^2$ scales--well above the transition from hadron to quark-gluon degrees of freedom. The EIC will add spectacular precision to pdf extraction at low $x$; estimates give a few percent uncertainty for sea distributions and on the order of $10\%$ for nuclear glue. Inclusive charm measurements, the so-called $F_2^{\rm charm}$, will also contribute to a large reduction in gluon PDF uncertainties at large $x$. 

The precision of the lepton probe, and high energies, will allow careful studies of light and heavy hadron production, in and out of the nucleus, in both light and heavy nuclei.  The production of jets and jet-hadron/jet-photon correlations in $e+A$ collisions will be feasible for the first time. Di-hadron (and in principle multi-hadron) correlations will be studied. Empirical knowledge of how struck light and heavy quarks (as well as heavy Onia), scatter, shower and hadronize in nuclear media will be greatly expanded relative to HERMES, Jlab6, COMPASS and Jlab12 experiments. 

One should note that by the time EIC rolls around, considerable knowledge on such ``cold QCD matter" effects will have been gleaned from the p+A programs at RHIC and LHC. However separating properties of the hadron/nuclear wavefunction from those of the scattering process, is not trivial to achieve;  in a number of instances, the complementarity between p+A and e+A collisions is essential for deeper understanding. This is not an instance of theorists splitting hairs. As a concrete illustration,  consider the case of diffractive processes where no colored exchanges are permitted between the fragmentation phase space of the projectile and target. At the EIC, as we shall discuss shortly, such ``rapidity gap" events are a large part of the cross-section.  An identical diffractive measurement in p+A collisions will have a much smaller cross-section, since colored spectators in the projectile proton will interact with the nucleus, thereby destroying the rapidity gap. 

The study of the sizes and distributions of such gaps offer an opportunity to understand dynamically how color singlet exchanges are constituted and maintained across vast stretches in rapidity space. It matters because  even though these exchanges are long known to dominate hadron-hadron cross-sections at high energies, embarrassingly, we still don't understand their dynamics in QCD. 

\subsection{\em Discovery and characterization of universal gluon matter?}

An exciting possibility at EIC is the unambiguous discovery and characterization of universal strongly correlated gluon matter predicted to occur in the high energy Regge-Gribov limit of QCD. In this limit, the occupancy of gluons for all momentum modes up to a saturation scale $Q_S$ is the maximum possible in QCD. Because the occupancy is very large, the correspondence principle tells us that the physics of saturated gluons is classical physics. However because non-linear gluon self-interactions are important, the classical fields in QCD carry far more structure than those in QED. The scale $Q_S$ separating this classical regime of saturated fields--from the regime described more conventionally in terms of parton distributions-- is  an emergent dynamical scale in QCD that grows with increasing energy and atomic number. 

There are a number of features of saturated gluon matter that have the potential to radically transform our understanding of the strong interactions and may perhaps also hold lessons for strongly correlated systems in other fields.
A key feature is that even though the matter is strongly correlated, a number of its properties can be computed in weak coupling. This is remarkable because typically strongly correlated systems are also strongly coupled and there is no small parameter that one can utilize to perform computations of the properties of such systems. These are non-perturbative computations in the conventional sense; the results cannot be expressed as a simple analytical expansion in the small parameter, the coupling constant. This is to be contrasted with the dynamics of confinement and chiral symmetry breaking which control the dynamics of QCD at large distances where the coupling may be large. The opening up of a novel strongly correlated window at high energies, and its interplay with strong coupling dynamics, offers a fascinating opportunity to use the former (under better theory control) to understand important features of the latter.

The properties of saturated gluon matter are universal in the Regge-Gribov limit. To be specific, at distance scales smaller than the nucleon size, gluon distributions and correlations are independent of whether the  target is a hadron or a nucleus. Observables can be expressed as dimensionless combinations of the external hard scale and the saturation scale; asymptotically, their dependence on rapidity, analogously to dynamical scaling exponents in the theory of critical phenomena, can be  expressed in terms of pure numbers. These numbers, some studies suggest, may be those that characterize the universality class of an integrable Heisenberg spin chain~\cite{Lipatov:1993yb}. It is unlikely that this asymptotics will be realized in experiments, even at the high energy LHeC machine. Nevertheless the pre-asymptotics may carry signatures of this universal dynamics. 

An effective theory of QCD in Regge-Gribov asymptotics is the Color Glass Condensate (CGC), a name that attempts to capture the stochastic features of gluodynamics in high occupancy states~\cite{Gelis:2010nm}. The content of the CGC can be formulated in terms of a coupled set of renormalization group equations for multi-parton correlators that add flesh to above words and make concrete testable predictions. Is the CGC the correct effective theory of the high energy limit of QCD or are there perhaps other descriptions that are more complete that incorporate strongly coupled dynamics as well ?

In my view, the strongest experimental tests of the CGC will be from diffractive studies at the EIC. Nucleons, and nuclei more so, resemble black discs that are opaque to the propagation of colored projectiles, with only the diffractive shadow of the incoming wave surviving the scattering. Indeed an amusing fact is that as one approaches this black disc limit, even a probe that's one hundredth the size of the nucleon will have a scattering cross-section that's comparable to those measured in nucleon-nucleon scattering. This is very different from the other asymptotic Bjorken limit of QCD; in these asymptotics, the nucleus is nearly transparent to the colored probe. Diffractive measurements at a collider, which will be performed off nuclei for the first time at EIC, are striking measurements. Some models predict that the diffractive cross-section will be about a quarter of the total cross-section. In plain language, this corresponds to a TeV energy electron scattering off a nucleus at rest-- and the nucleus remaining intact in one in 4 collisions--despite its binding energy per nucleon being a million times smaller than the energy of the electron! 

The CGC effective theory suggests that the right degrees of freedom in high energy QCD are so-called dipole and quadrupole correlators. These can be extracted in semi-inclusive measurements in both proton-nucleus and electron-nucleus collisions. A predicted softening of back-to back correlations with increasing centrality amongst particles produced in the deuteron fragmentation region of deuteron-nucleus scattering experiments at RHIC has been seen; the recently concluded proton-nucleus run at RHIC will help distinguish the CGC based explanations from alternative descriptions. The corresponding measurements at the EIC will provide definitive tests of this framework. 

Another set of measurements in proton-proton and proton-nucleus collisions that have generated much excitement are the so-called ridge correlations. These are two particle correlations that are collimated in their relative azimuthal angle (around the beam axis) but are long range in their rapidity separation. Ridge correlations were seen previously in nucleus-nucleus collisions and have a straightforward interpretation in that case as arising from hydrodynamic flow. The situation is more complex in the smaller systems because the CGC initial state in the collision can also describe the ridge-like correlations in these small systems. The EIC will bring a unique perspective to the study of such correlations. Photo-production in electro-nucleus collisions, from vector dominance, is essentially a hadron-nucleus collision. Thus many of the measurements in high energy proton-nucleus collisions can also be performed at an EIC. One may be able to observe the ridge in very rare, high multiplicity photo-production events. However with increasing $Q^2$, we anticipate that the ridge correlations should die away. How and whether this occurs should enable us to isolate initial state contributions from those in the final state. These studies will provide a fresh angle on the fascinating questions on whether the ridge correlations reflect novel quantum correlations between the produced particles or whether we are observing collective flow of the world's smallest fluids.

\section{Further scientific opportunities with EIC}

The hitherto discussed topics address the primary elements of the science case for a future EIC. The history of such prognostications however promises that the science drivers after experiments are underway are often different from those projected aforehand. We will list below a few topics currently at the periphery of the EIC science discussion that have the potential to move front and center in the future. 

\begin{itemize}

\item {\em Testing fundamental symmetries at the EIC.}
The EIC has the potential to significantly impact studies of charged lepton flavor violation, precision measurements of the running of the electroweak angle and other searches of beyond-the-standard model physics. Very preliminary studies of these have been reported in the EIC whitepaper, and further study is warranted. These measurements are very luminosity hungry with requirements of over 100 fb$^{-1}$ essential to improve on current and anticipated measurements. The potential of EIC in BSM physics will depend a great deal on what we learn from the LHC in the next decade. 

\item {\em EIC and the structure of nuclei.}
Collider kinematics will enable exploration of the nuclear fragmentation region with precision high energy probes. With the tagging of nuclear spectators, and the wide array of light and heavy nuclei that will be available, novel studies of nucleon interactions (in particular short-range nucleon-nucleon interactions) will become feasible. The possible existence of exotic color octet--color octet configurations in nuclei can be investigated. 

\item {\em Hadron spectroscopy.}
The search for novel  glueballs, hybrids, tetra- and penta-quark color singlet states is flourishing with numerous searches ongoing at electron-positron colliders, the LHC, experiments at COMPASS and with the commissioning and operation of the new GlueX detector at Jlab. The discovery of so-called XYZ states and the recent announcement of a pentaquark candidate by LHCb have piqued interest. What can the EIC contribute 10 years hence? Jlab will contribute significantly to the search for light quark hybrids with exotic quark numbers. With the EIC, such studies can be extended to charm and bottom hybrids predicted in lattice computations. While complementarity with existing facilities is assured, EIC's discovery potential in this area will only become clearer in the next several years. 
\end{itemize}
\section{Acknowledgements}
This work was supported under DOE Contract No.  de-sc0012704. I am grateful to my BNL experimental colleagues Elke Aschenauer and Thomas Ullrich, as well as Rolf Ent from Jlab, and Abhay Deshpande of Stony Brook University, who have taught me some of the EIC lore. The support of my BNL theory colleagues is always forthcoming and greatly appreciated. I would also like to thank Marco Stratmann, Steve Vigdor and Werner Vogelsang for their very useful comments on the manuscript.

\end{document}